\documentclass[preprint,amsmath,amssymb,amsfonts,aps,superscriptaddress,nofootinbib]{revtex4}
\usepackage{ifsym}
\usepackage{MnSymbol}
\usepackage[usenames,dvipsnames]{xcolor}
\usepackage{bm}
\usepackage{hyperref}

\def\be{\begin{equation}}
\def\ee{\end{equation}}
\def\ba{\begin{eqnarray}}
\def\ea{\end{eqnarray}}

\def\nn{\nonumber}

\def\a{\alpha}
\def\b{\beta}

\def\m{\mu}
\def\n{\nu}

\begin{document}
%\preprint{APS/123-QED}

\title{Spontaneous Breaking of (2+1)-dimensional Lorentz Symmetry by an antisymmetric tensor}
\author{Jia-Hui Huang}
\email{huangjh@m.scnu.edu.cn}
\affiliation{Guangdong Provincial Key Laboratory of Quantum Engineering and Quantum Materials,
School of Physics and Telecommunication Engineering,
South China Normal University, Guangzhou 510006,China}
%\affiliation{Guangdong Provincial Key Laboratory of Quantum Engineering and Quantum Materials}
\author{Guang-Zhou Guo}
\affiliation{Guangdong Provincial Key Laboratory of Quantum Engineering and Quantum Materials,
School of Physics and Telecommunication Engineering,
South China Normal University, Guangzhou 510006,China}
\affiliation{Center for theoretical physics, College of Physical Science and Technology,
Sichuan University,
Chengdu, 610064, China}
\author{Hao-Yu Xie}\author{ Qi-Shan Liu}\author{ Fang-Qing Deng}
\affiliation{Guangdong Provincial Key Laboratory of Quantum Engineering and Quantum Materials,
School of Physics and Telecommunication Engineering,
South China Normal University, Guangzhou 510006,China}

\begin{abstract}
One kind of spontaneous (2+1)-dimensional Lorentz symmetry breaking is discussed. The symmetry breaking pattern is $SO(2,1)\rightarrow SO(1,1)$. Using the coset construction formalism, we derive the Goldstone covariant derivative and the associated covariant gauge field. Finally, the two-derivative low-energy effective action of the Nambu-Goldstone bosons is obtained.
\end{abstract}
\maketitle

\section{Introduction}
Lorentz symmetry plays an important role in modern physics. For example, it is the basic symmetry of special relativity and the standard model of particle physics. To date, the experimental constraints on Lorentz violation are still stringent\cite{Kostelecky:2008ts,Flambaum:2016dwc,Kostelecky:2016pyx}.

However, many efforts have been made to investigate Lorentz-violating models. A theoretical motivation is that quantum gravity is believed to violate Lorentz symmetry. For example, in string theory, the effective theory of D-branes with background Neveu-Schwarz--Neveu-Schwarz \textit{B} field is described by noncommutative field theory, which violate Lorentz symmetry \cite{Seiberg:1999vs,Higashijima:2001sq}. Spontaneous Lorentz symmetry breaking (SLSB) also has attracted attention in some phenomenological models which propose interesting explanations for gauge bosons. In the Abelian Nambu model\cite{Nambu:1968qk,Azatov:2005wv,Escobar:2015gia},  photon is emergent as the Nambu-Goldstone (NG) boson of SLSB and this model has been proved to be equivalent to the standard QED at tree\cite{Nambu:1968qk}, one-loop\cite{Azatov:2005wv}, and all loop orders in perturbation theory\cite{Escobar:2015gia}. Non-abelian gauge bosons can also be emergent as NG bosons of SLSB \cite{Chkareuli:2007da,Escobar:2015eoa}.

Another interesting application of SLSB is for condensed matter systems. The long-wavelength collective excitations of certain systems can be described by the NG bosons arising from spontaneously broken symmetries\cite{Jackiw,son02,son05,nicolis06,nicolis12,Nicolis:2013lma,Nicolis:2015,Delacretaz:2014jka}. This idea is applied to study low-energy dynamics of superfluid in \cite{son02} and is generalized to perfect fluids and solids\cite{nicolis06,nicolis12,Nicolis:2013lma}. In \cite{Nicolis:2015}, the authors classify different condensed matter systems in terms of the spacetime and internal symmetries which are spontaneously broken.

For spontaneous internal symmetry breaking, according to Goldstone theorem, there is one massless NG boson corresponding to each broken generator. The effective action of NG bosons can be obtained in coset construction formalism. This model-independent formalism  for internal symmetry breaking was developed by Coleman, Callan, Wess and Zumino (CCWZ)\cite{Coleman:1969sm,Callan:1969sn} and
was further generalized to spacetime symmetry breaking cases \cite{Volkov,Ogievetsky,ArmendarizPicon:2010mz}. Compared with internal symmetry breaking, there are two main differences for spontaneous spacetime symmetry breaking. One difference is the number of NG bosons may be less than the number of broken generators, which is known as inverse Higgs effect\cite{Low:2001bw,Goon:2014ika}. Another difference comes from the coset construction formalism. Since translations act nonlinearly on coordinates, the translation generators should be treated as "broken" generators, regardless of whether it is truly broken or not.

Generally, Lorentz symmetry is spontaneously broken by nonzero vacuum expectation values (VEVs) of vector fields or tensor fields. And, an antisymmetric 2-tensor is often used to trigger SLSB in literatures. The VEV of the antisymmetric 2-tensor can be a constant background\cite{Yokoi:2000en,Hernaski:2016dyk} or can be developed dynamically\cite{Higashijima:2001sq,Hosotani:1993dg,Altschul:2009ae,Hosotani:1994sc}.

In this paper, we focus on SLSB in (2+1)-dimensional flat spacetime. The breaking is triggered by the VEV of an antisymmetric 2-tensor, $F_{\m\n}$, which has three independent components. $F_{12}$ is the magnetic component and $F_{01}$, $F_{02}$ are the electric components. In Ref.\cite{Yokoi:2000en}, the author considered SLSB triggered by the nonzero VEV of magnetic component $F_{12}$ and derived the effective action of the NG bosons. Here we will study SLSB triggered by  the nonzero VEV of an electric component, say $F_{01}$. The symmetry breaking pattern is $SO(2,1)\rightarrow SO(1,1)$ and the common NG-boson counting rule applies. Using coset construction, we derive the two-derivative low-energy effective action of the NG bosons, which is the main result of the paper.

This paper is organized as follows. In Section II, we briefly review the coset construction for spacetime symmetry breaking. In section III, the spontaneous breaking of (2+1)-dimensional Lorentz symmetry is studied. The effective action for the NG bosons is finally derived. Section IV is devoted to the conclusion.
\section{coset construction}
\subsection{Internal symmetry breaking}
For spontaneous breaking of internal symmetry , the general formalism to construct the effective action of NG bosons was developed by Coleman, Callan, Wess and Zumino\cite{Callan:1969sn,Coleman:1969sm}. An internal symmetry group $G$  is spontaneously broken to its unbroken subgroup $H$. The group generators should satisfy the following relations
\begin{equation}
\begin{split}
\left[T_i,T_j\right]&=if_{ijk}T_k ,\\
\left[T_i,X_a\right]&=if_{iab}X_b ,\\
\left[X_a,X_b\right]&=if_{abc}X_c+if_{abi}T_i ,\\
\end{split}
\end{equation}
where $T_i$ denote the unbroken subgroup generators in $H$ and $X_a$ denote the broken generators in $G/H$. If $f_{abc}$ in the third commutation vanish, $G/H$ is called symmetric coset space. The NG bosons $\xi_a$ are used to parameterize the coset space through
\begin{equation}
U\left(\xi\left(x\right)\right)=e^{i\xi_a\left(x\right)X_a}.
\end{equation}
This parametrization defines a nonlinear realization of the group $G$. Considering the action of an element $g\in G$ on  $U$, we have the following  transformation
\begin{equation}\label{gTrans}
gU\left(\xi\left(x\right)\right)=U\left(\xi'\left(x\right)\right)h\left(\xi\left(x\right),g\right),
\end{equation}
where $h\left(\xi\left(x\right),g\right)\in H$. From the  parameterized coset element $U$, one can define the Goldstone covariant derivatives $D_{\mu a}$ and the associated gauge fields $A_{\mu i}$ from Cartan-Maurer one-form
\begin{equation}
U^{-1}dU=D+A=idx^\mu\left(D_{\mu}\xi_a X_a+A_{\mu i}T_i\right).
\end{equation}
Under the general group transformation \eqref{gTrans}, it is straightforward to derive the corresponding transformation rules for $D$ and $A$,
 \ba
 D\rightarrow h D h^{-1},~~~ A\rightarrow h A h^{-1}+h d h^{-1}.
 \ea
 The associated gauge field $A_\mu$ can be used to define covariant derivative for other matter field $\psi$ which furnish a linear representation of $H$,
$ D_\mu\psi=\partial_\mu \psi+A_\mu\psi$.
 With the covariant quantities $D_\mu$ and $D_\mu\psi$, one can construct the invariant effective action. For example, the two-derivative effective action of  NG bosons is
 \ba
 S_2(\xi)=\int d^Dx\frac{1}{2}\mathrm{Tr}(D_\mu D^\mu).
 \ea

\subsection{Spacetime symmetry breaking}
The coset construction formalism for spacetime symmetry breaking was developed in \cite{Ogievetsky,Volkov}. Here we just consider Poincare symmetry breaking. One special thing is that we should take the translation generators as broken generators, even actually they are not broken. For simplicity, the same symbols are used for other broken and unbroken generators as the internal case. The parameterization of the coset space is
\begin{equation}
U\left(x,\xi\left(x\right)\right)=e^{i x^\mu P_\mu}e^{i\xi_a\left(x\right)X_a}.
\end{equation}
Under the action $g$ of $G$, the transformation of $U$ is
\begin{equation}\label{gTransST}
gU\left(x,\xi\left(x\right)\right)=U\left(x', \xi'\left(x'\right)\right)h\left(\xi\left(x\right),g\right),
\end{equation}
where $h\left(\xi\left(x\right),g\right)\in H$. The Maurer-Cartan 1-form in this case is
\begin{equation}
U^{-1}dU=E+D+A=i dx^\mu\left(e^{\ \a}_{\mu}P_\a+D_{\mu}\xi_a X_a+A_{\mu i}T_i\right).
\end{equation}
Compared with the internal case, extra terms $e_\m^{\ \a}$ appear and play the role of  vielbeins. Under the group transformation \eqref{gTransST}, we have
\ba
E\rightarrow h E h^{-1},~~
D\rightarrow h D h^{-1},~~
A\rightarrow h A h^{-1}+hdh^{-1}.
\ea
With these quantities, we can define the covariant derivatives of the NG fields $\xi_a$ and other matter field $\psi$ which belongs to a linear representation of $H$,
\begin{equation}
\nabla_\a\xi_a=e_{\ \a}^{\nu}D_\nu\xi_a,~~\nabla_\a\psi=e_{\ \a}^{\nu}(\partial_\nu\psi+A_\nu\psi),
\end{equation}
where $e_{\ \a}^{\nu}$ is the inverse of the vielbein, satisfying  $e_{\ \a}^{\nu}e_{\nu}^{\ \b}=\delta_\a^\b$. In the above definition, vielbein plays important role. Then the general $G$-invariant effective action is given by
\begin{equation}
S(\xi,\psi)=\int \mathrm{det}\left(e_{\mu}^{\ \a}\right)d^Dx \mathcal{L}(\nabla_\a\xi_a,\nabla_\a\psi, \psi).
\end{equation}

\section{Spontaneous (2+1)-dimensional Lorentz symmetry breaking}
In this section, we will consider the spontaneous breaking of (2+1)-dimensional Lorentz symmetry. The metric is chosen as $\eta_{\mu\nu}=\left(+1, -1, -1\right)$. Because the speciality of translations in spacetime symmetry breaking, we should start with the full Poincare symmetry. The commutation relations of the Poincare generators are
\ba\label{duiyi}
\left[P^\mu,P^\nu\right]=0,~~
\left[J^\mu,P^\nu\right]=-i\epsilon^{\mu\nu\rho}P_\rho,~~
\left[J^\mu,J^\nu\right]=-i\epsilon^{\mu\nu\rho}J_\rho,
\ea
where $J^\mu\equiv\frac{1}{2}\epsilon^{\mu\nu\rho}M_{\nu\rho}$ and $M_{\nu\rho}$ are the Lorentz generators.

In \cite{Yokoi:2000en}, the Lorentz symmetry is broken by an antisymmetric 2-tensor $F_{\mu\nu}$ with nonzero constant vacuum value: $<F_{12}>\neq 0,~<F_{0i}>=0$. Two boost generators $J^1,J^2$ are broken and space rotation generator $J^0$ is unbroken. The symmetry breaking pattern is $so(2,1)\rightarrow so(2)$.

Here we will consider the complementary case. The nonzero constant vacuum value is taken as $<F_{01}>\neq 0,~<F_{02}>=<F_{12}>=0$. Then it is straightforward that the generators $J^0,J^1$ are broken and $J^2$ is unbroken. The symmetry breaking pattern is $so(2,1)\rightarrow so(1,1)$. In fact, the detail of the symmetry breaking is not important and the coset construction formalism is only related with the symmetry breaking pattern.

One should keep in mind that the translations should be taken as broken generators. The element of the coset  $SO(2,1)/SO(1,1)$ is parameterized by NG fields $\phi_0, \phi_1$ as follows,
\begin{equation}\label{U}
U\left(x,\phi_a\left(x\right)\right)=e^{ix^\mu P_\mu}e^{i(\phi_0(x)J^0+\phi_1(x)J^1)}
=e^{i\left(x^+P_++x^-P_-+x^2P_2\right)}e^{i\left(\phi_+ J^++\phi_- J^-\right)},
\end{equation}
where
\begin{equation}\label{redefine}
\begin{split}
x^+=x^0+x^1&,\quad  x^-=x^0-x^1,\\
\phi_+=\phi_0\left(x\right)+\phi_1\left(x\right)&,\quad \phi_-=\phi_0\left(x\right)-\phi_1\left(x\right),\\
P_+=\frac{1}{2}\left(P_0+P_1\right)&,\quad P_-=\frac{1}{2}\left(P_0-P_1\right),\\
J^+=\frac{1}{2}\left(J^0+J^1\right)&,\quad J^-=\frac{1}{2}\left(J^0-J^1\right).\\
\end{split}
\end{equation}
Here we use the light-cone coordinates for convenience. Under the unbroken boost transformation,  the transformation rules of NG fields and space-time coordinates are
\ba\label{UnBroTran}
x^+\rightarrow e^{-\eta}x^+ ,~x^-\rightarrow e^\eta x^-,~\phi_{+}\rightarrow e^\eta\phi_+,~\phi_{-}\rightarrow e^{-\eta}\phi_-,
\ea
where $\eta$ is the rapidity.

From \eqref{U}, we have the following Cartan-Maurer one-form
\begin{equation}\label{cmone}
U^{-1}dU=idx^\mu\left(e^{\ \a}_{\mu}P_\a+D_\mu\phi_+ J^++D_\mu\phi_- J^-+A_\mu\left(x\right)J^2\right),~~(\m,\a=+,-,2).
\end{equation}
After some calculation (details are in Appendix),
we can derive the vielbein
\begin{equation}\label{vielbein}
(e_\mu^{\ \a})=\begin{pmatrix}
  \frac{\cosh\phi+1}{2} & \frac{\phi^2_-}{\phi^2}\frac{\cosh\phi-1}{2} & -\frac{\phi_-}{\phi}\sinh\phi\\
  \frac{\phi^2_+}{\phi^2}\frac{\cosh\phi-1}{2} & \frac{\cosh\phi+1}{2} & \frac{\phi_+}{\phi}\sinh\phi\\
  \frac{\phi_+}{\phi}\frac{\sinh\phi}{2} & -\frac{\phi_-}{\phi}\frac{\sinh\phi}{2} & \cosh\phi\\
  \end{pmatrix},
\end{equation}
and other components in the Cartan-Maurer one-form
\ba
D_\mu\phi_+&=&\partial_\mu\phi_++\partial_\mu\left[\frac{\phi_+}{\phi}\left(\sinh\phi-\phi\right)\right],\\\label{Dphi}
D_\mu\phi_-&=&\partial_\mu\phi_-+\partial_\mu\left[\frac{\phi_-}{\phi}\left(\sinh\phi-\phi\right)\right],\\\label{A}
A_\mu&=&\frac{\partial_\mu\phi_+\phi_--\phi_+\partial_\mu\phi_-}{2\phi^2}\left(\cosh\phi-1\right).
\ea
where $\phi^2=-\phi_+\phi_-=\phi^2_1-\phi^2_0$. Comment is needed here. From the definition, $\phi^2$ is not positive definite. So one may doubt the definition of $\phi$ in the above  equations \eqref{vielbein}-\eqref{A}. In fact, when we expand the above equations in $\phi$, $\phi$ always appears in powers of $\phi^2$. The above equations are well defined.

The inverse of the vielbein \eqref{vielbein} can be derived straightforwardly,
\begin{equation}\label{invVie}
(e^\nu_{\ \b})=\begin{pmatrix}
  \frac{\cosh\phi+1}{2} & \frac{\phi^2_-}{\phi^2}\frac{\cosh\phi-1}{2} & \frac{\phi_-}{\phi}\sinh\phi\\
  \frac{\phi^2_+}{\phi^2}\frac{\cosh\phi-1}{2} & \frac{\cosh\phi+1}{2} & -\frac{\phi_+}{\phi}\sinh\phi\\
  -\frac{\phi_+}{\phi}\frac{\sinh\phi}{2} & \frac{\phi_-}{\phi}\frac{\sinh\phi}{2} & \cosh\phi\\
  \end{pmatrix}.
\end{equation}
From the discussion of previous section, the Goldstone covariant derivatives  are defined as
\begin{equation}
\nabla_\a\phi_i=e^\nu_{\ \a}D_\nu\phi_i,\quad \left(i=+,-\right),
\end{equation}
which can be written explicitly as follows,
\ba\nonumber\label{GoldCov1}
\nabla_+\phi_i&=&\frac{\cosh\phi+1}{2}\left[\partial_+\phi_i+\partial_+\frac{\phi_i}{\phi}\left(\sinh\phi-\phi\right)\right]
+\frac{\phi^2_+}{\phi^2}\frac{\cosh\phi-1}{2}\left[\partial_-\phi_i+\partial_-\frac{\phi_i}{\phi}\left(\sinh\phi-\phi\right)\right]\\
&&+\left(-\frac{\phi_+}{\phi}\frac{\sinh\phi}{2}\right)\left[\partial_2\phi_i+\partial_2\frac{\phi_i}{\phi}\left(\sinh\phi-\phi\right)\right],\\\nonumber
\label{GoldCov2}
\nabla_-\phi_i&=&\frac{\phi^2_-}{\phi^2}\frac{\cosh\phi-1}{2}\left[\partial_+\phi_i+\partial_+\frac{\phi_i}{\phi}\left(\sinh\phi-\phi\right)\right]
+\frac{\cosh\phi+1}{2}\left[\partial_-\phi_i+\partial_-\frac{\phi_i}{\phi}\left(\sinh\phi-\phi\right)\right]\\
&&+\frac{\phi_-}{\phi}\frac{\sinh\phi}{2}\left[\partial_2\phi_i+\partial_2\frac{\phi_i}{\phi}\left(\sinh\phi-\phi\right)\right],\\\nonumber
\label{GoldCov3}
\nabla_2\phi_i&=&\frac{\phi_-}{\phi}\sinh\phi\left[\partial_+\phi_i+\partial_+\frac{\phi_i}{\phi}\left(\sinh\phi-\phi\right)\right]
+\left(-\frac{\phi_+}{\phi}\sinh\phi\right)\left[\partial_-\phi_i+\partial_-\frac{\phi_i}{\phi}\left(\sinh\phi-\phi\right)\right],\\
&&+\cosh\phi\left[\partial_2\phi_i+\partial_2\frac{\phi_i}{\phi}\left(\sinh\phi-\phi\right)\right]
\qquad \qquad \qquad \left(i=+,-\right).
\ea

Using these Goldstone covariant derivatives, we can construct the effective action  of NG bosons, which is invariant under full group $G$.
Under the transformation \eqref{UnBroTran}, the covariant Goldstone derivatives of NG bosons transform as follows
 \ba
\nabla_+\phi_+ &\rightarrow& e^{2\eta}\nabla_+\phi_+,~~~\nabla_+\phi_- \rightarrow \nabla_+\phi_-,\\
\nabla_-\phi_+ &\rightarrow& \nabla_-\phi_+,~~~~~~\nabla_-\phi_-\rightarrow e^{-2\eta}\nabla_-\phi_-,\\
\nabla_2\phi_+ &\rightarrow& e^\eta\nabla_2\phi_+,~~~~\nabla_{2}\phi_{-}\rightarrow e^{-\eta}\nabla_2\phi_-.
 \ea
To construct the effective action, we also need the invariant integral measure. In our model, this measure is given by
\ba
\det(e_\m^{\ \a})d^3x=d^3x.
\ea
 Then the two-derivative effective action of NG bosons is expressed as
\ba\nn\label{EfAc}
S_{2}(\phi_+,\phi_-)&=&\int d^3x \left\{\right.a_1\nabla_+\phi_-+a_2\nabla_-\phi_+
+a_3\nabla_+\phi_-\nabla_-\phi_++a_4\left(\nabla_+\phi_-\right)^2\\&&+a_5\left(\nabla_-\phi_+\right)^2
+a_6\nabla_+\phi_+\nabla_-\phi_-+a_7\nabla_2\phi_+\nabla_2\phi_-\},
\ea
where $a_i(i=1,...,7)$ are seven arbitrary real parameters.

When  considering the coupling of NG bosons and other matter fields $\psi$, we need the associated covariant gauge field, which are defined as
 \ba\label{gaugefield}
 A_\a=e^\nu_{\ \a} A_\nu.
 \ea
Then the covariant derivative of the matter field $\psi$  is
 \ba
\nabla_\a\psi= e^\nu_{\ \a}\partial_\nu\psi+A_\a\psi.
 \ea
The effective action for NG bosons and $\psi$ is
 \ba
 S_2(\phi,\psi)=\int d^3x \mathcal{L}(\nabla_\a\phi,\nabla_\a\psi,\psi),
 \ea
where $\mathcal{L}$ is $H$-invariant. This effective action is invariant under the transformation of full group $G$.

\section{Conclusion}
We investigate the spontaneous breaking of (2+1)-dimensional Lorentz symmetry. The breaking pattern is $SO(2,1)\rightarrow SO(1,1)$, which is caused by the nonzero vacuum of the electric component $F_{01}$ of an antisymmetric 2-tensor. There are two NG bosons $\phi_0$ and $\phi_1$, which correspond to the low-energy quantum excitations of $F_{02}$ and $F_{12}$ respectively.  The Goldstone covariant derivatives are derived in \eqref{GoldCov1}- \eqref{GoldCov3} and the associated covariant gauge field is in \eqref{gaugefield}. The two-derivative low-energy effective action of the two NG bosons is finally obtained in \eqref{EfAc}.

When the 2-tensor $F_{\mu\nu}$ is an antisymmetric gauge field, it has no physical degree of freedom in three dimensional spacetime, so the resulting massless NG bosons are unphysical. If the 2-tensor is field strength of a vector field, the massless NG bosons are physical. An example of this case can be found in \cite{Hosotani:1993dg}, where the author proposed a three-dimensional gauge model with a Chern-Simons term where the Lorentz symmetry was spontaneously broken by a dynamically generated  magnetic field.

An interesting problem for further study is to find  physical models where the Lorentz symmetry breaking pattern discussed by us occurs. Another interesting thing is to find  possible condensed matter systems which can be described by the (2+1)-dimensional Lorentz symmetry breaking caused by non-zero magnetic or electric fields.

{\textbf{Acknowledgements:\\}}
This work is supported by the Natural Science Foundation of Guangdong Province, No.2016A030313444.

\appendix*
\section{}
In the calculation of  the Cartan-Maurer one-form \eqref{cmone}, the Baker-Campbell-Hausdorff formula
 \ba
 e^X Y e^{-X}=Y+[X,Y]+\frac{1}{2!}[X,[X,Y]]+\cdots,
 \ea
and the following formula
\begin{equation}
e^{X}d e^{-X}=e^{X}(-dX) e^{-X}=-\sum_{n=0}^\infty\frac{1}{(n+1)!}\left[X,\left[X,\cdots,
\left[X,dX\right]\cdots\right]\right],
\end{equation}
are useful.
Using these formulas, the Cartan-Maurer one-form \eqref{cmone} reads
\ba\nn
U^{-1}dU&=&e^{-i\pi}(idx^\mu P_\mu) e^{i\pi}+e^{-i\pi}(id\pi)e^{i\pi}\\\nn
&=&idx^\mu\Bigg\{P_\mu+(-i)[\pi,P_\mu]+\frac{(-i)^2}{2!}[\pi,[\pi,P_\mu]]+\frac{(-i)^3}{3!}[\pi,[\pi,[\pi,P_\mu]]]+\cdots\\\label{example}
&&+\partial_\mu\pi+\frac{-i}{2!}
\left[\pi,\partial_\mu\pi\right]+\frac{\left(-i\right)^2}{3!}\left[\pi,\left[\pi,\partial_\mu\pi\right]\right]+\frac{\left(-i\right)^3}{4!}
\left[\pi,\left[\pi,\left[\pi,\partial_\mu\pi\right]\right]\right]+\cdots\Bigg\},
\ea
where $\pi=\phi_+J^++\phi_-J^-$ .

Next, we show the calculation of two examples in  \eqref{cmone}. The first is the element $e_+^{\ +}$ in the vielbein, which is the coefficient of $idx^+P_+$, and the second is the associate gauge field $A_\mu$, which is the coefficient of $dx^\mu J^2$. Form  \eqref{duiyi} and  \eqref{redefine}, it is straightforward to derive the following commutation relations,
\ba
\left[\pi,P_+\right]&=&\frac{i}{2}\phi_+ P_2,\\
\left[\pi,\left[\pi,P_+\right]\right]&=&\frac{i^2}{2}\left(\phi^2P_++\phi_+^2 P_-\right),~\\
\left[\pi,\left[\pi,\left[\pi,P_+\right]\right]\right]&=&\frac{i^3}{2}\phi^2\phi_+P_2,\\
\left[\pi,\partial_\mu\pi\right]&=&\frac{i\left(\phi_-\partial_\mu\phi_+-\phi_+\partial_\mu\phi_-\right)}{2}J^2,~~\\
\left[\pi,\left[\pi,\partial_\mu\pi\right]\right]&=&\frac{\left(\phi_-\partial_\mu\phi_+-\phi_+\partial_\mu\phi_-\right)}{2}\left(\phi_+J^+-\phi_-J_-\right),\\
\left[\pi,\left[\pi,\left[\pi,\partial_\mu\pi\right]\right]\right]&=&\frac{i\left(\phi_-\partial_\mu\phi_+-\phi_+\partial_\mu\phi_-\right)}{2}
\left(-\phi^2\right)J^2.
\ea
 Other higher order commutation relations can easily be derived. Plugging these relations into \eqref{example}, we can obtain the vielbein element and gauge filed
\ba
e_+^{\ +}&=&1+\frac{(-i)^2}{2!}\left(\frac{i^2}{2}\phi^2\right)+\frac{(-i)^4}{4!}\left(\frac{i^4}{2}\phi^4\right)+\cdots=\frac{\cosh\phi+1}{2},\\\nn
A_\mu&=&\frac{\left(\partial_\mu\phi_+\phi_--\phi_+\partial_\mu\phi_-\right)}{2}\left(\frac{-i}{2!}i+\frac{\left(-i\right)^3}{4!}i\left(-\phi^2\right)+\frac{\left(-i\right)^5}{6!}i\left(-\phi^2\right)^2+\cdots\right)\\
&=&\frac{\phi_-\partial_\mu\phi_+-\phi_+\partial_\mu\phi_-}{2\phi^2}\left(\cosh\phi-1\right).
\ea
Other terms in the  Cartan-Maurer one-form can be calculated similarly.

\end{document}